# BiosecurID: A Multimodal Biometric Database

J. Fierrez · J. Galbally · J. Ortega-Garcia · M. R. Freire · F. Alonso-Fernandez · D. Ramos · D. T. Toledano · J. Gonzalez-Rodriguez · J. A. Siguenza · J. Garrido-Salas · E. Anguiano · G. Gonzalez-de-Rivera · R. Ribalda · M. Faundez-Zanuy · J. A. Ortega · V. Cardeñoso-Payo · A. Viloria · C. E. Vivaracho · Q. I. Moro · J. J. Igarza · J. Sanchez · I. Hernaez · C. Orrite-Uruñuela · F. Martinez-Contreras · J. J. Gracia-Roche

This work has been supported by the Spanish MEC under project TIC2003-08382-C05-01. The authors J. G. and R. R. are supported by FPU Fellowships from Spanish MEC, the authors M. R. F. and F. A.-F. are supported by FPI Fellowships from CAM, and J. F. is supported by a Marie Curie Fellowship from the European Commission.

J. Fierrez · J. Galbally · J. Ortega-Garcia · M. R. Freire · F. Alonso-Fernandez · D. Ramos · D. T. Toledano · J. Gonzalez-Rodriguez · J. A. Siguenza · J. Garrido-Salas · E. Anguiano · G. Gonzalez-de-Rivera R. Ribalda
Universidad Autonoma de Madrid, EPS,
C/ Francisco Tomas y Valiente 11, 28049 Madrid, SPAIN
Tel.: +34-914973363
Fax: +34-914972235
E-mail: {julian.fierrez, javier.galbally, javier.ortega}@uam.es

M. Faundez-Zanuy
Escuela Universitaria Politecnica de Mataro,
Avda. Puig i Cadafalch 101-111, 08303 Mataro, Barcelona, SPAIN

J. A. Ortega
Universidad Politecnica de Cataluña, Esc. Univ. de Ing. Tec. Ind. de Terrassa,
C/ Colom 1, 08222 Terrassa, Barcelona, SPAIN

V. Cardeñoso-Payo · A. Viloria · C. E. Vivaracho · Q. I. Moro
Universidad de Valladolid, Edif. de Tecnicas de la Inf. y las Telecom.,
Campus Miguel Delibes s/n, 47011 Valladolid, SPAIN

J. J. Igarza · J. Sanchez · I. Hernaez
Universidad del Pais Vasco, Escuela Superior de Ingenieros,
C/ Alameda de Urquijo s/n, 48013 Bilbao, SPAIN

C. Orrite-Uruñuela · F. Martinez-Contreras · J. J. Gracia-Roche
Universidad de Zaragoza, Computer Vision Lab Group, Edif. Ada Byron,
C/ Maria de Luna 1, 50015 Zaragoza, SPAIN

**Abstract** A new multimodal biometric database, acquired in the framework of the BiosecurID project, is presented together with the description of the acquisition setup and protocol. The database includes 8 unimodal biometric traits, namely: speech, iris, face (still images and videos of talking faces), handwritten signature and handwritten text (on-line dynamic signals and off-line scanned images), fingerprints (acquired with two different sensors), hand (palmprint and contour-geometry) and keystroking. The database comprises 400 subjects and presents features such as: realistic acquisition scenario, balanced gender and population distributions, availability of information about particular demographic groups (age, gender, handedness), acquisition of replay attacks for speech and keystroking, skilled forgeries for signatures, and compatibility with other existing databases. All these characteristics make it very useful in research and development of unimodal and multimodal biometric systems.[1]

**Keywords** Biometrics . Multimodal . Database . Face . Speech . Iris . Fingerprint . Palmprint . Hand geometry . Keystroking . Signature . Handwriting

## 1 Introduction

Authentication methods based on biometric technology [1], which guarantees that the means of identification cannot be stolen, lost or forgotten, are being increasingly demanded in security environments and applications like access control and electronic transactions. Big efforts have been undertaken in the biometric community to increase the reliability of security systems by combining existing unimodal biometric matchers in order to implement multimodal authentication systems [2, 3]. Nevertheless, in real-world circumstances, error rates achieved with state-of-the-art technology have slowed down their generalized application. In order to overcome the difference in performance between laboratory experiments and practical implementations, there is an urgent need for the collection of realistic multimodal biometric data which permits to infer valid results from controlled experimental conditions to the final application.

In the present contribution we describe the BiosecurID Biometric Multimodal Database acquired within the BiosecurID project, and conducted by a consortium of 6 Spanish Universities: Universidad Autonoma de Madrid (UAM), Universidad Politecnica de Madrid (UPM), Universidad Politecnica de Cataluña (UPC, Campus of Terrasa and Campus of Mataro), Universidad de Zaragoza (UniZar), Universidad de Valladolid (UVA), and Universidad del Pais Vasco (UPV/EHU). The main objective of the project was the acquisition of a realistic multimodal and multisession database, statistically representative of the potential users of biometric applications, and large enough in order to infer valid results from its usage.

---

[1] The distribution details of the BiosecurID database are available at http://atvs.ii.uam.es/databases.jsp

The creation of multimodal databases like BiosecurID alleviates one of the main problems in early research works on multimodal biometrics, which is the usage of "chimerical subjects". Due to the difficulty in obtaining large multimodal databases, some researchers have opted in their studies to combine different unimodal databases [4, 5], thus using "chimerical subjects" based on the assumption of independence between different biometric traits. This approach has been severely questioned in the literature [6], and is seen by many as a serious methodological flaw [7].

Although several real multimodal biometric databases are already available for research purposes, none of them can match the BiosecurID database in terms of number of subjects, number of biometric traits and number of temporally separated acquisition sessions. The data collected in the project are especially useful for the development and testing of automatic recognition systems due to some design characteristics such as: realistic acquisition scenario, balanced gender and population distributions, availability of information about particular demographic groups (age, gender, handedness, visual aids), acquisition of replay attacks (speech and keystroking) and skilled forgeries (signatures) in order to simulate attacking scenarios, and compatibility with other existing databases.

The paper is structured as follows. Related works on multimodal biometric databases are first summarized. This is followed by the description of the BiosecurID database, the acquisition environment, and a detailed explanation of the acquisition protocol. The validation and post-processing steps conducted in order to obtain the final release of the database are then described. The paper concludes after presenting some potential uses of the database.

## 2 Related works

Some significant examples of multimodal biometric databases, either completed and already available, or in process of completion are [7,8]:

- BIOMET [9] includes five different modalities: audio, face images (2D and 3D), hand images, fingerprint (captured with an optical and a capacitive sensor), and signature. The database was acquired in three temporally separated sessions (8 months between the first and the last one) and comprises 91 subjects who completed the whole process.
- MyIDEA [10] includes face, audio, fingerprints, signature, handwriting and hand geometry. Two synchronized recordings were also performed: face-voice and writing-voice. The general specifications of the database are: target of 104 subjects, different quality sensors, various realistic acquisition scenarios with different levels of control, organization of the recordings to allow an open-set of experimental scenarios, and compatibility with other existing databases such as BANCA [11].
- BIOSEC [12] was acquired under the BioSec Integrated Project, and comprises fingerprint images acquired with three different sensors, frontal face images from a webcam, iris images from an iris sensor, and voice utterances (captured both with a webcam and a close-talk headset). The baseline corpus described in [12] comprised 200 subjects with 2 acquisition sessions per subject. The extended version of the BIOSEC database comprises 250 subjects with 4 sessions per subject (about 1 month between sessions).
- BIOSECURE [13] is one of the results of the Biosecure Network of Excellence. This new database was designed and is being constructed by the same research groups that led MyIDEA, BIOSEC, and BiosecurID. The rationale behind BIOSECURE database is to extend those previous databases in scale (ca. 1000 subjects in BIOSECURE, being a group of them the same acquired in BIOSEC and BiosecurID), under various realistic acquisition conditions. In particular, the database considers three acquisition scenarios [13], namely:

  - *Internet Dataset*. The following data is captured in an unsupervised setup through the Internet: voice and face (still images and talking faces).
  - *Desktop Dataset*. The acquisition setup represents an office-like scenario and the acquisition of the following biometrics is conducted with human supervision: voice, fingerprints (two sensors), face (still images and talking faces), iris, signature (genuine and skilled forgeries) and hand.
  - *Desktop Dataset*. The acquisition of the following data is conducted using two mobile devices (a PDA and a Ultra-Mobile PC): signature (genuine and skilled forgeries), fingerprints (sweep sensor), voice, and face (images and video).

  All datasets include 2 sessions, with the biggest dataset (internet) comprising around 1,000 subjects, and the other two about 700 users. Around 400 of these subjects are common to the whole database.

Some other multimodal databases are the MCYT database [14] including fingerprints and signatures of 330 subjects, the MBioID database [15] acquired to study the use of biometrics in Identity Documents (2D and 3D face, fingerprint, iris, signature and speech), the BANCA database [11] comprising face and voice recordings of 208 subjects, and the new multibiometric, multidevice and multilingual M3 database [16], which includes face, speech (in Cantonese, Putonghua and English) and fingerprint traits captured using three different devices (desktop PC, pocket PC and 3G mobile phone) of 32 subjects. Other examples are FRGC [17], XM2VTS [18], SmartKom [19] or BT-DAVID [20].

In Table 1 a summary of the most relevant features of existing multimodal biometric databases is presented. In order to present all the information in a compact manner, both palmprint and palm geometry are considered as Hand trait, and on-line and off-line signature as Signature trait. In case of a different number of subjects in each acquisition session (as is the case of the BIOMET database) the number of subjects common to all the sessions is presented.

**Table 1** Summary of the most relevant features of existing multimodal biometric databases. The nomenclature followed is: # stands for *number of*, 2Fa for Face 2D, 3Fa for Face 3D, Fp for Fingerprint, Ha for Hand, Hw for Handwriting, Ir for Iris, Ks for Keystroking, Sg for Signature, and Sp for Speech.

|  | #Subjects | #Sessions | #Traits | 2Fa | 3Fa | Fp | Ha | Hw | Ir | Ks | Sg | Sp |
|---|---|---|---|---|---|---|---|---|---|---|---|---|
| **BiosecurID** | 400 | 4 | 8 | X |  | X | X | X | X | X | X | X |
| **BIOSECURE** | Int. ca. 1,000 | 2 | 2 | X |  |  |  |  |  |  |  | X |
|  | PC ca. 700 | 2 | 6 | X |  | X | X |  | X |  | X | X |
|  | Mob. ca. 700 | 2 | 4 | X |  |  | X |  |  |  | X | X |
| **BIOSEC** | 250 | 4 | 4 | X |  | X |  |  | X |  |  | X |
| **MyIDEA** | ca. 104 | 3 | 6 | X |  | X | X | X |  |  | X | X |
| **BIOMET** | 91 | 3 | 6 | X | X | X | X |  |  |  | X | X |
| **MCYT** | 330 | 1 | 2 |  |  | X |  |  |  |  | X |  |
| **BANCA** | 208 | 12 | 2 | X |  |  |  |  |  |  |  | X |
| **MBioID** | ca. 120 | 2 | 5 | X |  | X |  |  | X |  | X | X |
| **M3** | 32 | 3 | 3 | X |  | X |  |  |  |  |  | X |
| **XM2VTS** | 295 | 4 | 2 | X |  |  |  |  |  |  |  | X |
| **SmartKom** | 96 | 172 | 4 |  |  |  | X | X |  |  | X | X |
| **FRGC** | 741 | Variable | 2 | X | X |  |  |  |  |  |  |  |

**Table 2** Statistics of the BiosecurID database.

|  | **BiosecurID DB. 400 subjects** |
|---|---|
| **Gender Distribution** | 54% (Male) / 46% (Female) |
| **Age Distribution** | 42% (18–25) / 22% (25–35) / 16% (35–45) / 20% (>45) |
| **Handedness** | 93% (Righthanded) / 7% (Lefthanded) |
| **Manual Workers** | 7% (Yes) / 93% (No) |
| **Vision Aids** | 66% (None) / 27% (Glasses) / 7% (Contact lenses) |

## 3 The BiosecurID database

The BiosecurID database was collected at 6 different sites in an office-like uncontrolled environment simulating a realistic scenario. One of its unique characteristics is the large number and variety of biometric traits considered (see Table 1 for a comparison with other biometric databases):

– Biometric traits: speech, iris, face (photographs and talking faces videos), signature and handwriting (on-line and off-line), fingerprints, hand (palmprint and contour-geometry), and keystroking.

Another two remarkable characteristics of the database are (see Table 1):

– Number of subjects: a total of 400 subjects. The number of subjects acquired per site, and the distribution in the database is: UAM 130 (IDs 1–130), UPC Mataro 40 (IDs 131–170), UPC Terrasa 35 (IDs 171–205),

UVA 77 (IDs 206–282), UPV/EHU 52 (IDs 283–334), UniZar 66 (IDs 335–400).
- Number of sessions: 4 sessions distributed in a 4 month time span. Thus, three different levels of temporal variability are taken into account: 1) within the same session (the samples of a given biometric trait are not acquired consecutively), 2) within weeks (between two consecutive sessions), and 3) within months (between non-consecutive sessions). This is specially relevant in traits such as face, speech, handwriting or signature which present a significant variation through time.

Another design principle in the BiosecurID database was to have balanced demographic groups representing the potential users of biometric applications in Spain. Thus, all sites were asked to acquire 30% of the subjects between 18 and 25 years of age, 20% between 25 and 35, 20% between 35 and 45, and the remaining 30% of the subjects above 45 years of age. The actual statistics included in Table 2 show the difficulty in recruiting old donors in comparison to youngsters, who were more easily available and predisposed. The gender distribution was forced to be balanced and only a 10% difference was permitted between male and female sets. The ethnicity of most subjects correspond to white/caucasian Spaniards.

All relevant non-biometric data of each subject was stored in an independent file (available with the biometric samples) so that experiments regarding specific demographic groups can be easily carried out. The available information in these files includes: age, gender, handedness, manual worker (yes/no), and vision aids (glasses, contact lenses, none). The "manual worker" group includes all subjects having eroded fingerprints, as identified by the contributors themselves when asked about their daily tasks (e.g., drivers, farm laborers, etc.) In Table 2 the actual statistics of the BiosecurID database are shown.

## 4 Acquisition environment

Each of the 6 acquisition sites prepared an acquisition kiosk following some general indications about the environmental conditions, regarding illumination (neutral lighting with no preponderant focuses), noise (indoor conditions with no excessive background noise), and pose of the contributor (sitting in a non-revolving chair). This relaxed environmental conditions allow a desirable variability between the samples acquired in the different sites (e.g., background in facial images) which simulates the changing working conditions of a real-world biometric application. In Fig. 1 we show the acquisition kiosk prepared in one of the sites, together with some of the devices used in the acquisition.

During the acquisition procedure a human operator gave the necessary instructions to the contributors so that the acquisition protocol was followed. In spite of this guidance, and of the usage of a specifically designed acquisition software, some human and software errors occurred. In order to ensure that the BiosecurID database complies with the acquisition protocol, all biometric samples were manually verified by a human expert who either corrected or discarded the non-valid data. The guidelines followed in the validation process are further described in Sect. 6.

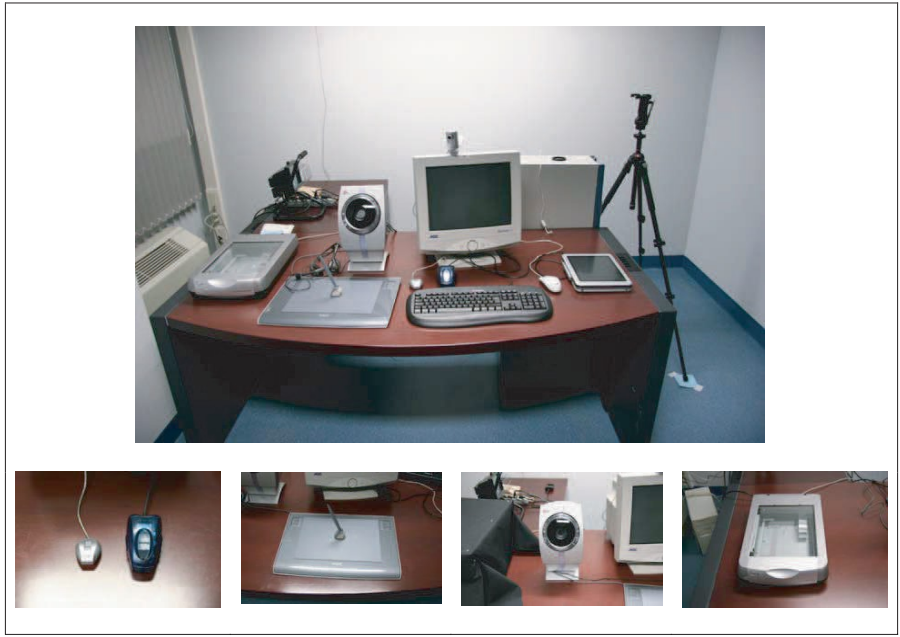

**Fig. 1** Example setup used in the acquisition of the BiosecurID database.

In Table 3 we show a list with all the devices used in the database acquisition and its most relevant features. All of them were connected to a standard PC in which the acquisition software tool, specifically designed following the database protocol, was run. This program centralized the functioning and launching of all the devices, as well as the naming and storage of the captured samples and management of the database, thus minimizing eventual acquisition errors. This tool also provided a common working interface for all the acquisition sites, in order to make the acquisition process faster, more reliable and homogeneous. In Fig. 2 screen captures of the different acquisition modules are shown.

## 5 Acquisition protocol

The biometric data along with the personal information captured are personal data and thus have to be protected according to the directives of the country where the responsible institution (or *controller* )[2] of the acquisition and management of the data is located, which for BiosecurID is Universidad Autonoma de Madrid in Spain. At the start of the first session a consent form was signed by each subject in which the subjects were properly informed about how the personal information will be used, that these data will only be transmitted to other institutions for research purposes and for a limited period of time, and that they have the right to access their data in order to correct, or delete it. The acquisition procedure started only once this consent form was fully understood and signed by the subject. Other requirements of the Spanish data protection authority are[3]: the controller must keep track of the licenses granted for the use of the database, the controller must adhere to certain security measures to protect the privacy of the subjects, and the database has to be entered in a national register of personal data files.

---

[2] Directive 95/96/EC of the European Parliament and the Council of 24 October 1995.

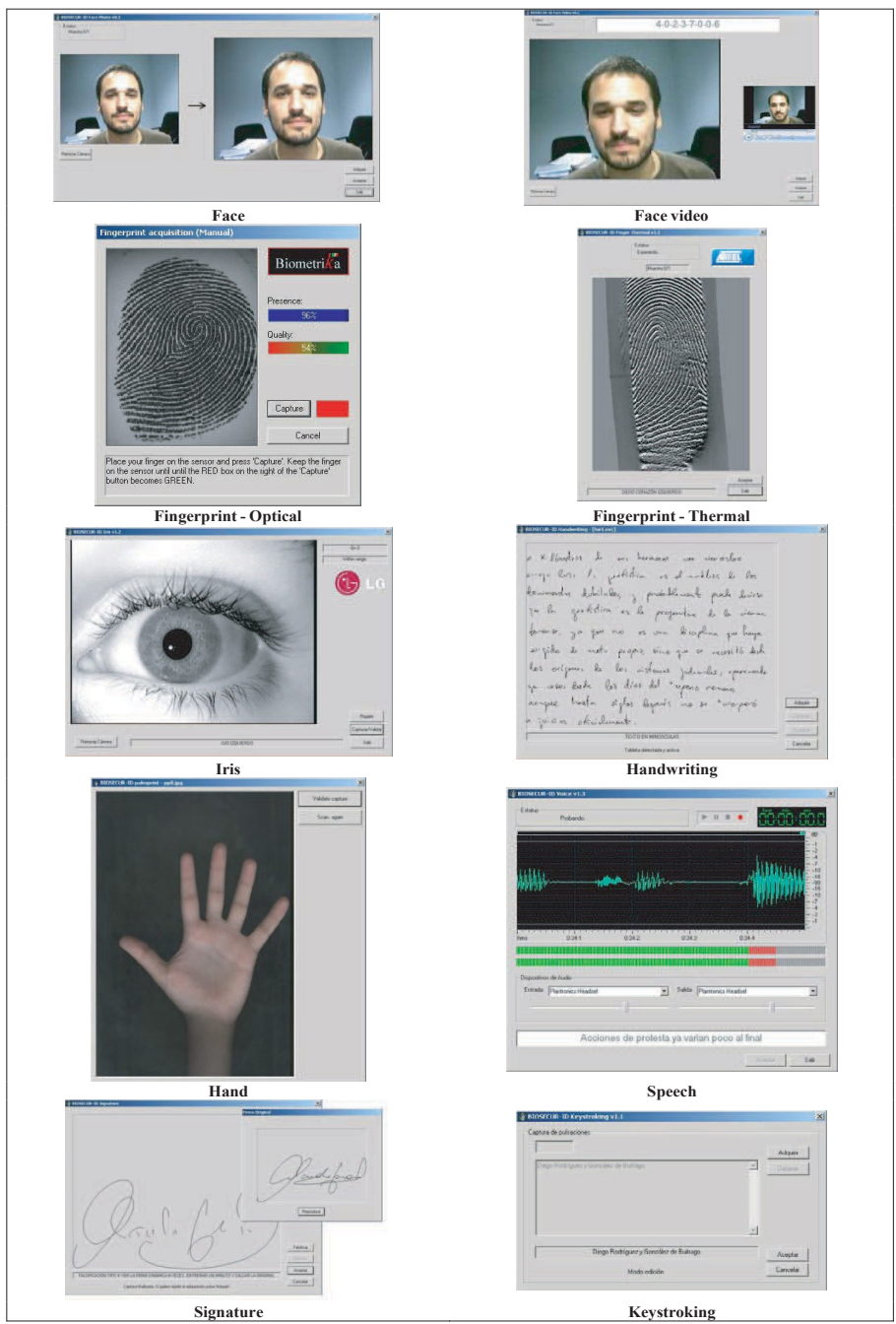

**Fig. 2** Screen captures of the different BiosecurID software acquisition modules.

**Table 3** Acquisition devices used for the BiosecurID database.

| Modality | Model | Main Features |
|---|---|---|
| Speech | Plantronics DSP 400 | Noise cancelling. 10Hz - 10KHz. |
| Fingerprints | Biometrika FX2000 | Optical. 569 dpi. |
| | | Capture area: 13.2 × 24.9 mm. |
| | | Image size: 400 × 560 pixels. |
| Fingerprints | Yubee (Atmel sensor) | Thermal Sweeping. 500 dpi. |
| | | Capture area: 13.9 × 0.5 mm. |
| | | Image size: 280 × 8 pixels. |
| Iris | LG Iris Access EOU 3000 | CCD. Infrared illumin. |
| | | Image size: 640 × 480 pixels. |
| Hand | Scanner EPSON Perfection 4990 | 4800 × 9600 dpi. 48 bits color depth. |
| | | Capturing area: 216 × 297 mm. |
| Face | Philips ToUcam Pro II | CCD. Illumin. 1 lux. |
| | | Image size: 640 × 480 pixels. |
| Writing/Signature | Wacom Intuos3 A4/Inking pen | 5080 dpi. 1024 pressure levels. |
| | | Accuracy: +/- 0.25 mm. |
| Keystroking | Labtec Standard Keyboard SE | Standard. |

In Table 4 we summarize the data samples of each biometric trait captured for every subject, namely:

– SPEECH: 10 short sentences in Spanish (the ones used in the Ahumada database [21], the same 10 for each subject) distributed along the four sessions (4 + 2 + 2 + 2) recorded at 44KHz stereo with 16 bits (PCM with no compression). In addition to the short sentences, 4 utterances of a subject-specific PIN of 8 digits were also recorded, and a utterance of other 3 subjects' PINs to simulate replay attacks in which an impostor has access to the number of a client. The forged subjects in each session were $n\,3S + 2$, $n\,3S + 1$, and $n\,3S$, where $n$ is the ID number inside the database of the current subject, and $S = 1, 2, 3, 4$ is the session number. The 8 digits were always pronounced digit-by-digit in a single continuous and fluent utterance.

---

[3] Ley Organica 15/99 (B.O.E. 14/12/1999).

**Table 4** Biometric data for each subject in the BiosecurID database (400 subjects in total).

| Modality | Samples | # Samples | Storage space (Mb) |
|---|---|---|---|
| Speech | 10 short sentences | 10 | 6.1 |
| | 4 × 4 PIN genuine | 16 | 15.2 |
| | 3 × 4 PIN imitations | 12 | 11.4 |
| | | **38** | **32.7** |
| Fingerprints | 4 × 4 × 4 optical | 64 | 10.2 |
| | 4 × 4 × 4 thermal | 64 | 12.3 |
| | | **128** | **22.5** |
| Iris | 2 × 4 × 4 | **32** | **9.4** |
| Hand | 2 × 4 × 4 | **32** | **11.6** |
| Face | 4 × 4 still faces | 16 | 14.1 |
| | 1 × 4 talking faces videos | 4 | 68.7 |
| | | **20** | **82.8** |
| Writing | 1 × 4 lower-case text | 4 | 2.4 |
| | 1 × 4 upper-case words | 4 | 1.2 |
| | 1 × 4 number sequence | 4 | 0.1 |
| | | **12** | **3.7** |
| Signature | 4 × 4 genuine signatures | 16 | 0.6 |
| | 3 × 4 skilled forgeries | 12 | 0.4 |
| | | **28** | **1.0** |
| Keystroking | 4 × 4 genuine name | 16 | 0.02 |
| | 3 × 4 skilled forgeries | 12 | 0.01 |
| | | **28** | **0.03** |

- FINGERPRINTS: 4 samples (BMP format with no compression) with 2 different sensors (see Table 3) of the index and middle fingers of both hands, interleaving fingers between consecutive acquisitions in order to achieve intravariability among images of the same fingerprint.
- IRIS: 4 samples (BMP with no compression) of each iris, changing eyes between consecutive captures. Glasses are removed for the acquisition, while the use of contact lenses is saved in the non-biometric data file.
- HAND: 4 images (JPG format) of each hand, alternating hands between consecutive acquisitions. The scanner used in the acquisition was isolated from external illumination using a box with just a little slot to insert the hand, and covered with a black opaque cloth.
- FACE: 4 frontal images (BMP not compressed), with no specific background conditions (except that no moving objects are permitted). One video sequence of five seconds saying the 8 digit PIN corresponding to the captured subject. Both the audio (PCM 8 bit) and video (29 frames per second) are captured with the webcam (see Table 3). No movement in the background is permitted.
- HANDWRITING: A Spanish text (the same for all subjects) handwritten in lower-case with no corrections or crossing outs permitted. The 10 digits, written separately and sequentially from 1 to 9 and last the 0. 16 Spanish separate words in upper-case. All the writing was captured using an inking pen so that both on-line dynamic signals (following the SVC format [22]) and off-line versions (scanned images) of the data are available. The lower-case text is collected in a different sheet of paper with no guiding lines, just a square highlighting the margins. The upper-case words and the number sequence were stored in a template-like page with boxes for each separate piece of writing.

- SIGNATURE: 4 genuine signatures per session (2 at the start and 2 at the end) and 1 forgery of each of the three precedent subjects (the same three in all the sessions). In order to consider an incremental level of skill in the forgeries, four different scenarios are considered, namely: *i*) the forger only sees the written signature once and tries to imitate it right away (session 1), *ii*) the subject sees the written signature and trains for a minute before making the forgery (session 2), *iii*) the subject is able to see the dynamics of the signing process 3 times, trains for a minute and then makes the forgery (session 3), and *iv*) the dynamics of the signature are shown as many times as the subject requests, he is allowed to train for a minute and then signs (session 4). Again both the on-line (SVC format) and off-line versions of the signature are captured using an inking pen. This trait is compatible with the publicly available MCYT database [14].
- KEYSTROKING: 4 case-insensitive repetitions of the subject's name and surname (2 in the middle of the session and two at the end) keystroked in a natural and continuous manner. No mistakes are permitted (i.e., pressing the backspace), if the subject gets it wrong, he is asked to start the sequence again. The names of 3 different subjects are also captured as forgeries (the same three subjects as in the speech PIN imitations), again with no mistakes permitted when keying the name. Samples are stored in plain text files with the total number of keystrokes in the first line, an event (SCAN code + D=press/U=release) and the miliseconds elapsed from the last event in the subsequent lines.

Imitations in the speech, signature and keystroking traits are carried out in a cyclical way, i.e., all the subjects imitate the previous subjects, and the first imitate the last subjects. Examples of typical images in the BiosecurID database are depicted in Fig. 3 (different traits corresponding to different random subjects). Voice utterances are shown as waveforms, both the dynamic signals and the scanned images are shown for the signatures and the handwritten text, while keystroking samples appear as bar plots of the sequence of keystrokes (press-down and inter-key times).

## 6 Validation process

Prior to the acquisition campaign and the validation process, the concepts *invalid sample* and *low quality sample* were defined so as to be certain of which biometric data were acceptable and which had to be rejected:

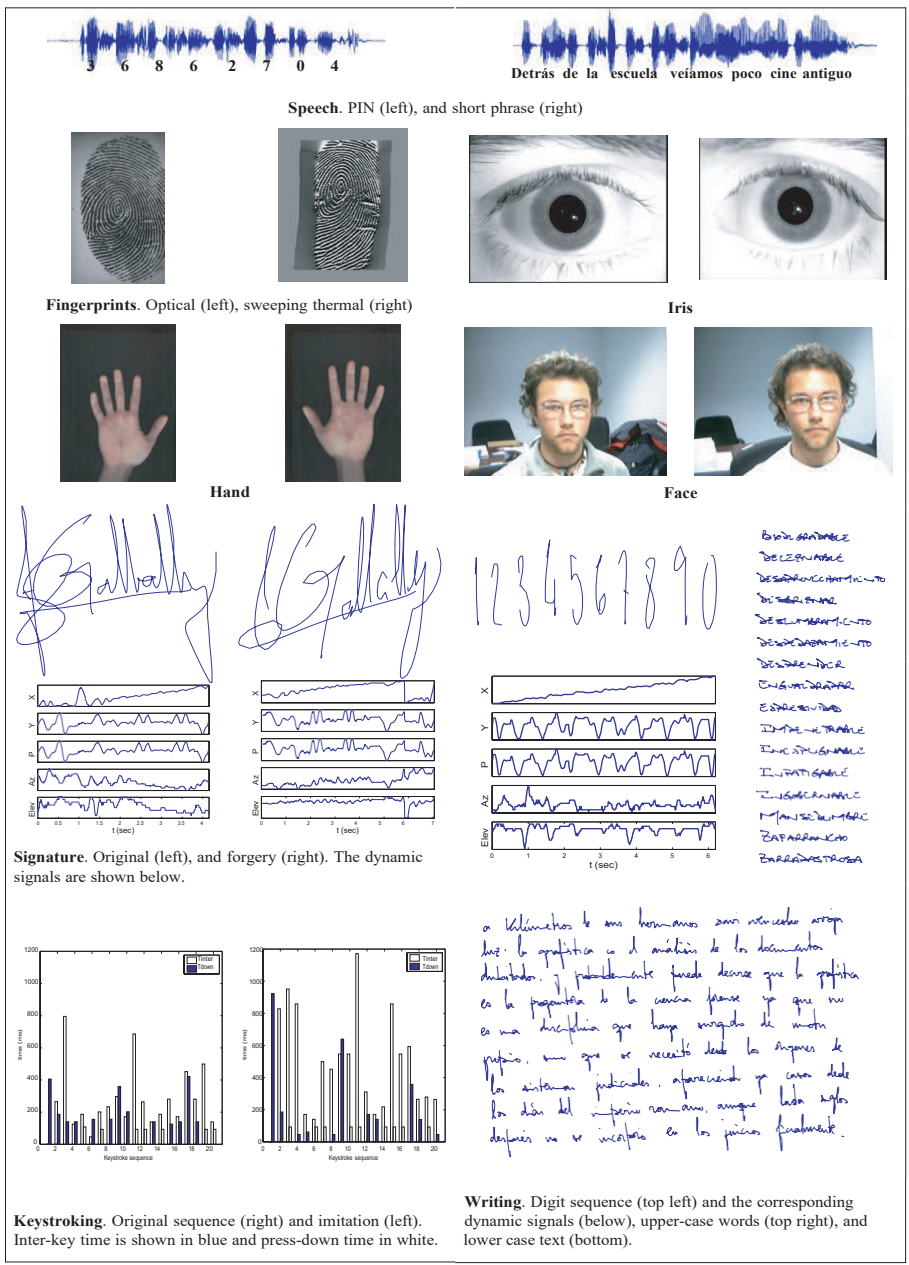

**Fig. 3** Samples of the different traits present in the BiosecurID database.

- Invalid sample. Is a sample that does not comply with the specifications given in the acquisition protocol (e.g., index finger labelled as middle, utterance of a wrong PIN, forgery of a wrong signature, etc.)
- Low quality sample. Is a sample that will typically perform badly on an automatic recognition system (e.g., very dry fingerprint image, wet fingerprint image, blurred iris image, voice utterance with high background noise, bad illumination in face images, side pose in face images, excessive pressure on a hand sample, etc.)

The main objective of the BiosecurID validation process was to reduce as much as possible the number of *invalid samples* within the database. The purpose of the procedure was in no case to reject *low quality samples*. Furthermore, the presence of low quality samples is a design feature of the database and a direct consequence of the non-controlled scenario where it was collected. Far from being a disadvantage, poor quality biometric data is an added value to the database as it is one of the key issues that real-world applications have to deal with. In this sense, BiosecurID is a suitable benchmark to evaluate how systems will perform in a realistic scenario. In Fig. 4 some of the typical biometric data that can be found in the BiosecurID database, and some selected low quality samples are shown.

The validation process of the biometric data in the BiosecurID database was carried out in two successive stages:

- Step 1. During the acquisition process a human supervisor aided by a specially designed acquisition software (see Sect. 4), validated one by one the captured samples, reacquiring those which were not compliant with the acquisition protocol.
- Step 2. Although the database was thus carefully collected, acquisition errors still happened. In order to ensure that the database fulfils all the acquisition specifications, all collected biometric samples were once again manually verified by a human expert who either completed the missing data, corrected invalid samples, or removed incomplete subjects.

The rules followed to either complete, correct or remove subjects from the database were the following:

- If a subject did not contribute to all the four sessions, then she was removed from the database.
- If a subject contributed to the four sessions, but in one or more of them an important part of her biometric data is missing or invalid (approximately more than 10% of all the genuine samples), then the subject was removed from the database.
- If a subject had a small number of missing or invalid genuine samples (approximately less than 10%), then those samples were copied from valid samples of the same subject in the same session. Therefore some identical genuine samples appear in some of the subjects in the BiosecurID database. These repeated samples can be easily filtered out prior to any experiment on BiosecurID, because they are the only identical genuine samples. These samples were included in BiosecurID in order to have a well structured and dimensioned database (which facilitates enormously the design of experiments, e.g., in case of user-dependent processing [23]), while keeping the subjects who provided large but incomplete sets of data.
- In the case of invalid or missing forgeries (PIN utterances, signature or keystroking), the expert verifying the database produced himself the missing or invalid samples.

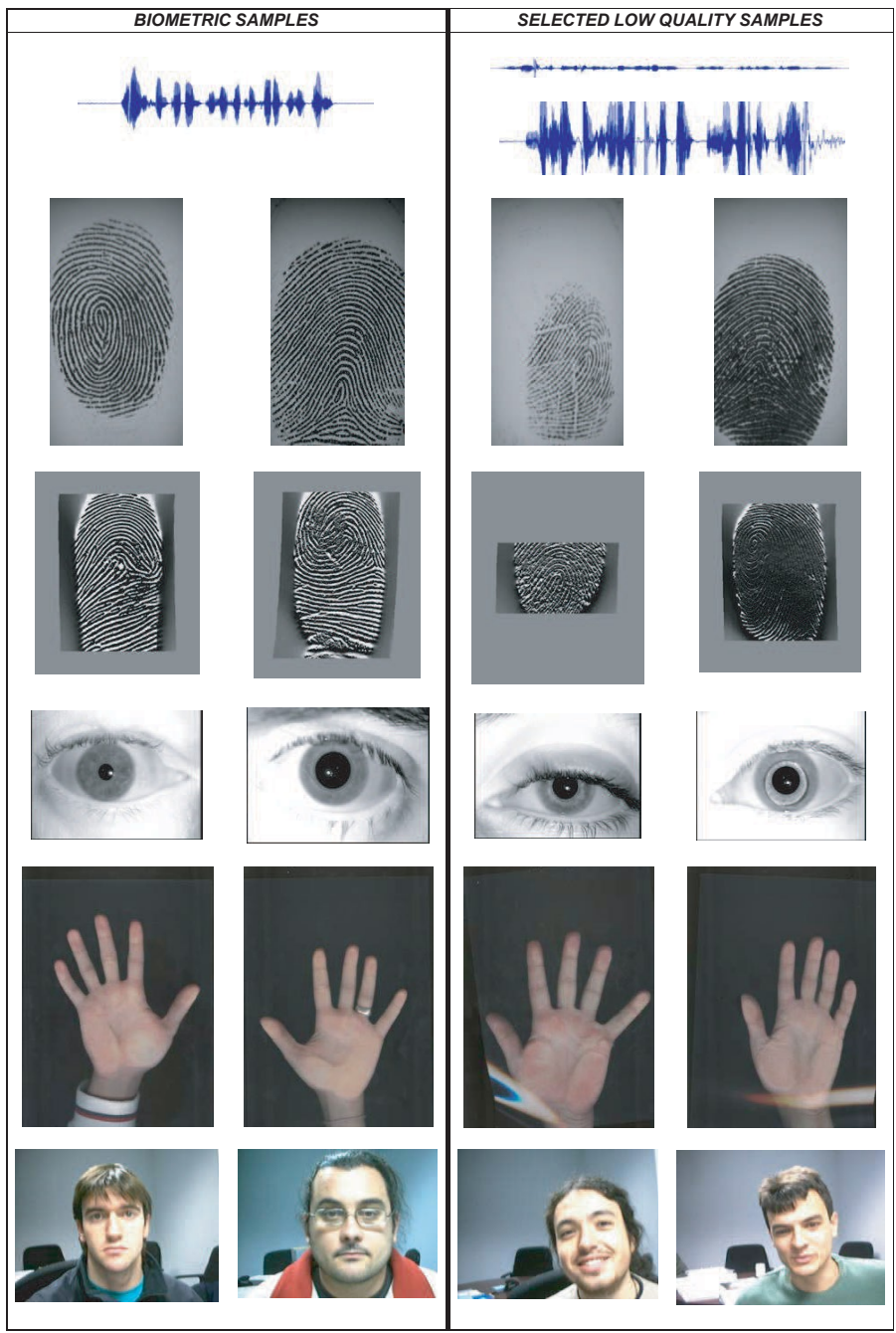

**Fig. 4** Typical biometric data (left), and selected low quality samples (right) that can be found in the BiosecurID database.

In spite of the careful acquisition process and the post editing efforts, some acquisition errors are very difficult to find and will only be detected through the usage of the database. Thus, after the initial release it is likely that future updated versions of the database will appear.

## 7 Compatibility with other databases

The devices and protocol used in the acquisition of some of the traits present in the BiosecurID database were chosen to be compatible with other existing databases, which enables new experimental setups combining various databases. Specifically, the BiosecurID database can be used in combination with:

- The BIOSEC database, which comprises 250 subjects. Both databases present compatible characteristics (sensors and protocol) in the following traits: optical/thermal fingerprints, face, speech and iris. This way, combining both datasets, a multimodal database of 650 subjects can be generated. Moreover, both databases (BIOSEC and BiosecurID) have 37 subjects in common, which allows to increase not only the number of subjects but also the number of sessions of the common subjects, thus permitting studies focused on long term biometric variability (about 1 year between BIOSEC and BiosecurID).
- The BIOSECURE Desktop and Mobile Datasets, with approximately 700 subjects in each dataset (ca. 400 subjects common to both of them). Similarly to the BIOSEC case, the BIOSECURE Desktop dataset is compatible with BiosecurID in optical/thermal fingerprints, iris, and signature. A group of 29 subjects participated both in BiosecurID and BIOSECURE and so again long term variability and interoperability studies can be performed upon them (about 1 year between BiosecurID and BIOSECURE).

Other multimodal databases such as MyIDEA (fingerprints, signature) or MCYT (signature) can also be combined with some portions of BiosecurID in order to increase the number of subjects as has been exposed with BIOSEC and BIOSECURE. However, in these cases no common subjects are available and so the number of sessions cannot be incremented. In Table 5 the main compatibilities of the BiosecurID database with other multimodal databases are summarized.

**Table 5** Summary of the main compatibilities of BiosecurID with other existing multimodal databases (total number of subjects in brackets). The abbreviations on the right column are the same used in Table 1.

|  | # Common subjects | Fa | Fp | Ha | Hw | Ir | Ks | Sg | Sp |
|---|---|---|---|---|---|---|---|---|---|
| **BIOSECURE Desktop** (ca. 700) | 29 |  | X |  |  | X |  | X |  |
| **BIOSECURE Mobile** (ca. 700) | 29 |  | X |  |  |  |  | X |  |
| **BIOSEC** (250) | 37 | X | X |  |  | X |  |  | X |
| **MyIDEA** (ca. 104) | 0 |  | X |  |  |  |  | X |  |
| **MCYT** (330) | 0 |  |  |  |  |  |  | X |  |

## 8 Potential uses of the database

Several potential uses of the database have already been pointed out throughout the paper. In this section some of the research lines that can be further developed upon this data set are summarized. It has to be emphasized that due to its unique characteristics in terms of size, acquisition environment and demographic distribution (age and gender), the BiosecurID database represents a good benchmark not only for the development of new algorithms, but also for testing existing approaches in the challenging acquisition conditions present in BiosecurID. Some of the possible uses of the database are (in brackets we indicate the database features that make possible the different studies):

- Research in any of the 8 available modalities or in multibiometric systems combining them (size, number of unimodal traits).
- Evaluation of the effect of time on the systems performance (multisession, compatible with other databases): 1) short term evaluation (samples within a session), 2) medium term evaluation (samples of different sessions), and 3) long term evaluation (considering the common subjects between BIOSEC/BIOSECURE and BiosecurID). This unique feature also enables research in biometric template adaptation and update [24].
- Studies on biometric sample quality and its effects on multibiometric systems (realistic uncontrolled acquisition scenario, verification process with low quality samples not discarded) [2].
- Research on the effect of the subjects age on the recognition rates, and its compensation (balanced age distribution) [25].
- Research and comparative studies of the systems performance depending on the gender of the subjects (balanced gender distribution) [26].
- Evaluation of sensor interoperability in those traits acquired with several devices (fingerprint, speech), and its effect on multibiometric systems (multidevice, number of traits) [27].
- Evaluation of potential attacks to unimodal, or multibiometric systems (size, number of unimodal traits) [28].

# 9 Conclusions

One of the main problems that can be found in the development, testing and evaluation of biometric recognition systems, is the lack of large public multimodal databases acquired under real working conditions. The importance of such databases should not be underestimated as it is often the failure of existing algorithms on new data sets, or simply the existence of new data sets, that drives research forward.

In the present contribution a short overview of existing multimodal biometric databases has been presented, together with a thorough description of the most relevant features of the new BiosecurID database, comprising speech, iris, face (photographs and talking faces videos), signature and handwriting (on-line dynamic signals and off-line scanned images), fingerprints (acquired with two different sensors), hand (palmprint and contour-geometry) and keystroking of 400 subjects, captured in 4 sessions along a 4 month time span.

The distribution details of the BiosecurID Multimodal Biometric Database are available at http://atvs.ii.uam.es/databases.jsp.

**Acknowledgements** The authors would like to thank Javier Garrido-Tomas, Borja Fernandez-Tomas and Daniel Hernandez-Lopez for their support during the acquisition campaign, and the valuable development work of J. Lopez-Peñalba and A. Posse.